# Optical bottle microresonators with axially-uniform eigenmode field distribution

## M. SUMETSKY

*Aston Institute of Photonic Technologies, Aston University, Birmingham B4 7ET, UK*
*m.sumetsky@aston.ac.uk*

**We show that the fundamental eigenmode of a shallow optical bottle microresonator (also called a SNAP microresonator) can be made exceptionally uniform along its axial length. The introduced microresonator has the effective radius variation resembling the profile of a bat with ears and wings. Remarkably, reduction of the axial size of this microresonator by cutting the wings does not alter the uniformity of its fundamental eigenmode. Being of general interest, our findings pave a way for improving the perceptibility of micro/nanoparticle sensing. These results also suggest a bottle microresonator suitable for accurate assembling of quantum emitters near the maximum of its eigenmode important in cavity quantum electrodynamics.**

Development of optical microresonator devices is the emerging area of photonics research with applications ranging from ultraprecise sensing [1-4] to quantum networking [5-7]. An optical microresonator can characterize the adjacent environment (liquid and gas [1-3], individual micro/nanoparticles and individual atoms and molecules [2-4]) through the measured variation of its spectrum with exceptional precision proportional to its Q-factor. Alternatively, high Q-factor microresonators can exchange photons with quantum emitters (quantum dots, vacancy centres, and individual atoms) and potentially can serve as single photon processing devices with applications in quantum computing and networking [5-7].

While methods for characterization of medium, which is uniform in the region of its interaction with the microresonator field, are well developed [1-4], sensing of objects which dimensions are comparable or much smaller than the microresonator dimensions (e.g., micro/nanoparticles, molecules, and atoms) is much more challenging. In fact, the same microparticle positioned at different locations of the microresonator will cause different shifts of its eigenmode proportional to the local intensity of the resonant field [8]. A possible solution to this problem is to create a microresonator having an eigenmode, which is *uniform* in the region of sensing. If a microparticle does not leave this region during the experiment then the observed shift of the eigenwavelength of such mode can be attributed to the change of the microparticle internal characteristics or orientation rather than its position. Thus, similar microparticles with similar orientation will cause the same variation of the resonance at such eigenmode. Complementary, the information about the motion of a microparticle can be extracted from monitoring the variations of spectra of other modes, which are not uniform in this region.

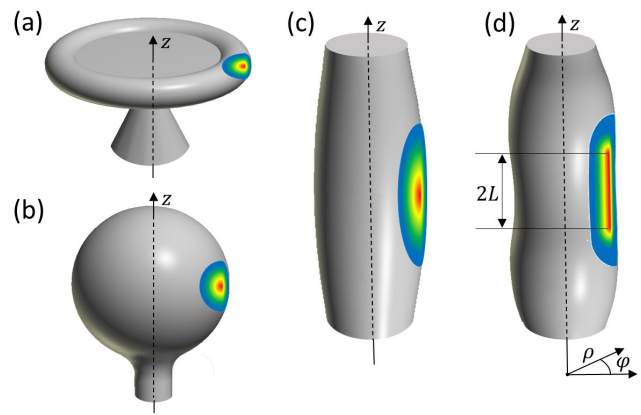

Fig. 1. Illustration of the fundamental WGM in (a) toroidal, (b) spherical, and (c) bottle microresonators, and in (d) a bottle microresonator with uniform WGM field distribution.

Optical microresonators having distended region with uniform eigenmode distribution near its maximum are also important in cavity quantum electrodynamics. In particular, in optical photonic crystal, ring, as well as toroidal and spherical microresonators, quantum emitters are positioned in the region close to their evanescent field maxima or antinodes and in maximum proximity to the resonator surface [9-15]. Typically, the characteristic size of the region where the field can be assumed constant has the size of a few tens of nanometers. While the problem of positioning accuracy along the direction towards the microresonator surface can be solved by placing quantum emitters directly onto the surface [9-13] and advancing the precision of optical tweezers [15], the problem of positioning of a quantum emitter close to the microresonator eigenmode maximum at the cavity surface requires a different approach. In particular, this problem can be solved with a microresonator which eigenmode is uniform along the extended surface area.

Figs. 1(a), (b), and (c) compare the characteristic fundamental whispering gallery mode (WGM) distribution in a toroidal, spherical and bottle microresonators. It is seen that the characteristic axial dimension of this mode is growing with the axial microresonator radius. Consequently, the vicinity of the WGM maximum, where the mode can be assumed uniform is growing as well. This vicinity can be enhanced in an elongated bottle microresonator (Fig. 1(c)). However, even for this resonator, the axial size of this vicinity is much smaller than the WGM axial dimension.

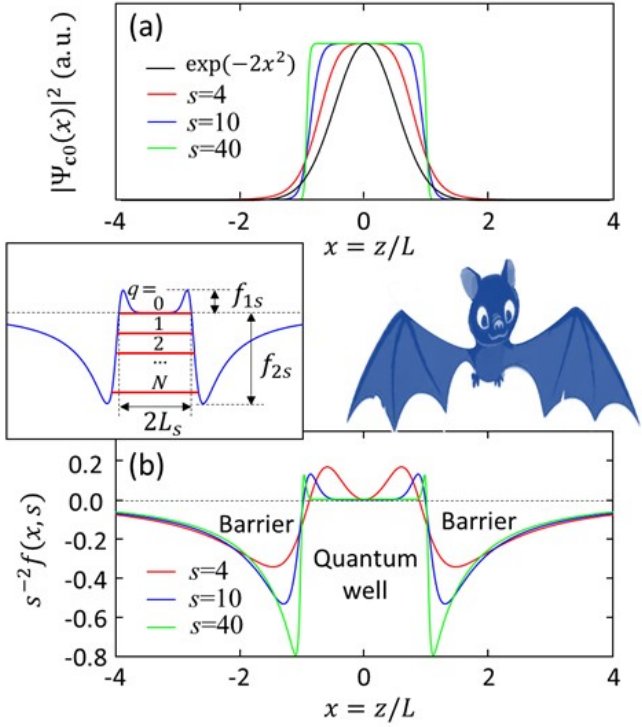

Fig. 2. (a) Intensity of the field distribution $|\Psi_{c0}(x)|^2$ defined by Eq. (4) for $s = 4, 10,$ and 40. (b) Dimensionless CWV $f(x,s)$ rescaled by the factor $s^{-2}$ for the same $s = 4, 10,$ and 40. Inset: The plot of dimensionless CWV $f(x,s)$ resembling the profile of a bat (drawn to the right of the inset) with parameters: ear height $f_{1s}$, wing width $f_{2s}$, and head width $2L_s$.

Here we consider shallow bottle microresonators with very small effective radius variation, which are also called SNAP resonators [16]. These resonators are usually fabricated from a uniform optical fiber by small deformation along its axial direction $z$. The eigenmodes of a SNAP bottle resonator are whispering gallery modes (WGMs) with small propagation constants $\beta_{cq}(z)$, which are determined at eigenwavelengths $\lambda_{cq}$ positioned close to a cutoff wavelength of the fiber $\lambda_c(z)$ as

$$\beta_{cq}(z) = 2^{3/2} \pi n_r \lambda_{cq}^{-3/2} (\lambda_c(z) - \lambda_{cq})^{1/2} \quad (1)$$

where the cutoff wavelength depends on the azimuthal and radial quantum numbers, $m$ and $p$, incorporated in $\mathbf{c} = (m, p)$ and eigenwavelengths $\lambda_{cq}$ also depend on the axial quantum number $q$. In Eq. (1), WGM losses are ignored. For an optical fiber with radius $r_0$ and refractive index $n_r$ having small variations $\Delta r(z)$ and $\Delta n_r(z)$, the cutoff wavelength variation (CWV) $\Delta \lambda_c(z)$ is determined from the equation $\Delta \lambda_c(z) / \lambda_{cq} = \Delta r_{eff}(z) / r_0 = \Delta r(z) / r_0 + \Delta n_r(z) / n_r$. This equation introduces the effective radius variation $\Delta r_{eff}(z)$ which combines variations of the radius and refractive index [16]. For a bottle microresonator with small CWV of our concern, the expression for WGMs is factorized in the cylindrical coordinates (Fig. 1(d)) as $E_{cq}(z, \rho, \varphi) = \exp(im\varphi) Q_c(\rho) \Psi_{cq}(z)$. Here $Q_c(\rho)$ satisfies the standard radial differential equation for an optical fiber [17] and the dependence on the axial coordinate is determined by the equation:

$$\frac{d^2 \Psi_{cq}(z)}{dz^2} + \beta_{cq}^2(z) \Psi_{cq}(z) = 0 \quad (2)$$

The problem addressed in this Letter is as follows. Suppose that we would like to create a SNAP microresonator which possesses an eigenmode with an axial quantum number $q$ having the predetermined axial field distribution $\Psi_{cq}(z)$. What should the corresponding CWV $\lambda_c(z)$ or effective radius variation $\Delta r_{eff}(z)$ be? Formally, the solution of this problem is straightforward. Indeed, we find from Eqs. (1) and (2):

$$\lambda_c(z) = \lambda_{cq} - \frac{\lambda_{cq}^3}{8\pi^2 n_r^2} \Psi_{cq}^{-1}(z) \frac{d^2 \Psi_{cq}(z)}{dz^2} \quad (3)$$

where $\lambda_{cq}$ is the wavelength eigenvalue of $\Psi_{cq}(z)$.

Here we consider the localized solutions of Eq. (2). For example, if the CWV $\lambda_c(z)$ is introduced along a segment of a uniform optical fiber then it is constant away from the microresonator, i.e., $\lambda_c(z) \to \lambda_c^{(\infty)}$ for $|z| \to \infty$. In this case, the boundary condition at $|z| \to \infty$ following from Eq. (3) is $\Psi_{cq}(z) \to \exp(-\kappa |z|)$ where $\kappa = 2^{3/2} \pi n_r \lambda_c^{-3/2} (\lambda_{cq} - \lambda_c^{(\infty)})^{1/2}$. Since the value of $\kappa$ is known in our problem and the value of $\lambda_c^{(\infty)}$ is determined by the properties of the uniform fiber, the latter equation determines the wavelength eigenvalue $\lambda_{cq}$. Obviously, localization of $\Psi_{cq}(z)$ inside the resonator requires $\lambda_{cq} > \lambda_c^{(\infty)}$ (i.e., real $\kappa$). In some cases, it is convenient to assume the boundary conditions that are different from that considered above. For example, for the Gaussian behavior of the field replicating the fundamental state of a harmonic oscillator in quantum mechanics [18] (a good approximation for the fundamental WGM localized in spherical and bottle microresonators), $\Psi_{c0}(z) = C \exp(-z^2 / L^2)$, Eq. (3) yields the parabolic dependence of the CWV, $\Delta \lambda_c(z) = (2\pi n_r L)^{-1} \lambda_c^3 [1 - 2(z/L)^2]$.

Can we design a bottle microresonator having an eigenmode which is constant along the axial direction $z$ like illustrated in Fig. 1(d)? The general answer to this question is given by Eq. (3). Indeed, the condition $\Psi_{cq}(z) = const$ immediately leads to $\lambda_c(z) \equiv \lambda_{cq}$, i.e., it requires that the propagation constant should be identically equal to zero, $\beta_c(z) \equiv 0$. The latter condition cannot be satisfied along the whole length of microresonator since $\Psi_{cq}(z)$ should vanish outside it. However, it can be satisfied with an exceptionally good accuracy along a finite segment of the resonator. As an example, we consider the behavior of the eigenmode of Eq. (2) in the form

$$\Psi_{c0}(z) = \left[1+\left(\frac{z}{L}\right)^s\right]^{-1} \quad (4)$$

Here $L$ is the characteristic axial width of the eigenmode and parameter $s$ is assumed an even integer. In Fig. 2(a), the eigenmode intensities $|\Psi_{c0}(z)|^2$ for $s = 4, 10,$ and $40$ (red, blue and green curves) are compared with the intensity of the fundamental eigenmode, $|\Psi_{c0}(z)|^2 = \exp(-2x^2/L^2)$ in a conventional microresonator with the parabolic variation of effective radius (black curve). It is seen that for large $s$ the distribution of $|\Psi_{c0}(z)|^2$ is close to uniform inside a finite interval with length $\sim L$ along axis $z$. From Eq. (3), we find the cutoff frequency variation:

$$\Delta\lambda_{c0}(z) = \lambda_c(z) - \lambda_{c0} = \frac{\lambda_c^3}{8\pi^2 n_r^2 L^2} f\left(\frac{z}{L}, s\right),$$
$$f(x, s) = sx^{s-2}(1+x^s)^{-2}\left[s - 1 - (1+s)x^s\right]. \quad (5)$$

where the dimensionless parameter

$$x = z/L. \quad (6)$$

From Eq. (5), function $f(x, s)$ determines the dimensionless profile of the CWV $\Delta\lambda_{cq}(z)$ and effective radius variation

$$\Delta r_{\text{eff}}(z) = \frac{\lambda_c^2 r_0}{8\pi^2 n_r^2 L^2} f\left(\frac{z}{L}, s\right). \quad (7)$$

The behavior of $f(x, s)$ is shown in Fig. 2(b) for $s = 4, 10,$ and $40$ corresponding to the distributions of eigenmodes $|\Psi_{c0}(z)|^2$ in Fig. 2(a). Similarity with the profile of a bat drawn above Fig. 2(b) ☺ suggests the name for this resonator: a bat bottle resonator (BBR). From Eqs. (1) and (5), we find the behavior of propagation constant along the BBR length:

$$\beta_{c0}(Lx) = \frac{1}{L} f(x, s)^{\frac{1}{2}}. \quad (8)$$

The eigenvalues of BBR are illustrated in the left hand side inset of Fig. 2(b) (red lines). In this inset, the BBR parameters – the ear height $f_{1s}$, the wing width $f_{2s}$, and the head width $2L_s$ – are introduced. From Eq. (5), the wavelength eigenvalue $\lambda_{c0}$ of $\Psi_{c0}(z)$ is equal to $\lambda_c^{(\infty)}$ since $f(x, s) \to 0$ for $|x| \to \infty$. In the middle of BBR where $f(x, s)$ is equal to zero, we have $\lambda_{c0} = \lambda_c^{(\infty)}$ as well. From Fig. 2(a) and (b), the interval with close to zero propagation constant corresponds to $\Psi_{c0}(z)$ close to constant. While the fundamental mode of BBR belongs to the discrete spectrum, all other modes have relatively small but finite widths due to the possibility of tunneling through potential barriers formed by the BBR wings (Fig. 2(b)). For $s \gg 1$, the total number of eigenmodes $N$ can be found under the assumption that the quantum well in Fig. 2(b) is rectangular with height $f_{2s} \approx s^2$ and width $2L_s \approx 2L$. Then in the semiclassical approximation we have $(f_{2s}/L)(4L) \approx 2\pi N$, i.e.,

$$N \approx 2s/\pi \quad (9)$$

and is independent on other BBR parameters.

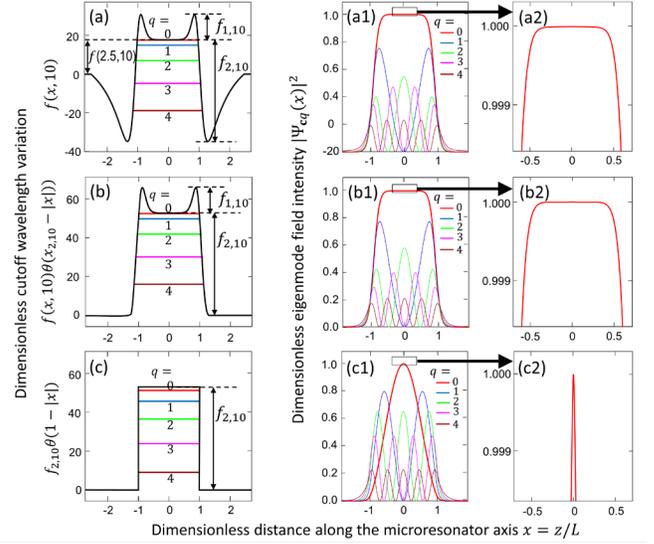

Fig. 3. The plots of dimensionless CWV (a) $f(x, 10)$, (b) the same as (a) but with cut wings $f(x, 10)\theta(x_{2,10} - |x|)$ where $x_{2,10} = 1.3$, and (c) with rectangular shape $f_{2,10}\theta(1 - |x|)$ where $f_{2,10} = 53.22$. (a1), (b1), and (c1) show the distribution of intensities of eigenmodes $|\Psi_{cq}(z)|^2$, $q = 0, 1, 2, 3, 4$, corresponding to microresonator profiles in (a), (b), and (c). (a2), (b2) and (c2) magnified behavior of fundamental modes $|\Psi_{c0}(z)|^2$ shown in (a1), (b1), and (c1) near their maxima.

Let us now consider a particular case of BBR having $s = 10$. The dimensionless plot of the CWV (or effective radius variation) of this microresonator is shown in Fig. 2(b) (blue curve) and Fig. 3(a). The characteristic dimensionless parameters in Fig. 3(a) are ear height $f_{1,10} = 13.21$, wing width $f_{2,10} = 53.22$, and the coordinates of maxima $x_{1,10} = 0.868$ and minima $x_{1,10} = 1.299$. This resonator has five eigenmodes $\Psi_{cq}(z)$ with quantum numbers $q = 0, 1, 2, 3$ and 4, which eigenvalues are shown in Fig. 3(a). These eigenvalues as well as the corresponding intensity distributions $|\Psi_{cq}(z)|^2$ shown in Fig. 3(a1) were found by numerical solution of Eq. (2). For this solution, the wings of BBR were partly cut (Fig. 3(a)). This approximation did not noticeably affect the behavior of BBR modes since the cut regions are inside the barriers far away from the area where eigenmodes $\Psi_{cq}(z)$ are localized. The latter fact suggests that the profile of the intensity of fundamental mode of this resonator $|\Psi_{c0}(z)|^2$ (bold red curve in Fig. 3(a1)) will not change if we go further and reduce the BBR size by cutting the BBR wings completely as shown in Fig. 3(b). The corresponding profile of $|\Psi_{c0}(z)|^2$ shown in Fig. 3(b1) (bold red curve) proves this assumption. Figs. 3(a3) and 3(b3) show the behavior of $|\Psi_{e0}(z)|^2$ for both models considered magnified near its maximum. Its seen that in both cases the uniformity of the fundamental eigenmode along the length $L$ of BBR is better than 0.1%. Remarkably, the analytical solution for the eigenmode determined by Eq. (4) does not deviate from numerical solutions in Figs. 3(a2) and (b2) within the precision of these figures.

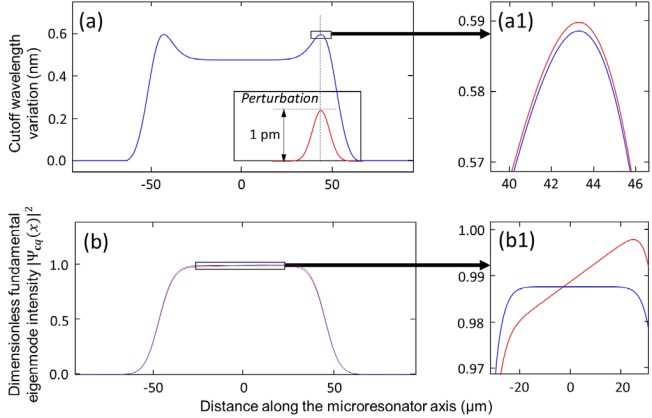

Fig. 4. (a) CWV for BBR with characteristic axial lengths $L = 50$ μm. The inset shows the introduced 1 pm high Gaussian perturbation of the profile of this BBR near its ear. (a1) Magnified comparison of perturbed and unperturbed BBRs near the ear. (b) The distribution of intensities of fundamental mode for unperturbed and perturbed BBRs. (b1) Plots shown in (b) magnified near their maxima.

Finally, we note that the experimental realization of the BBR described here is possible under two major conditions: (i) sufficient original uniformity of an optical fiber and (ii) sufficient fabrication precision of the microresonator introduced at the surface of this fiber. Remarkably, the available optical fibers have the appropriate uniformity with the CWV less and, presumably, much less than 1 pm along the lengths ~ 100 μm [19]. We estimate the feasibility of the second condition by considering the BBR model with characteristic length $L = 50$ μm at characteristic cutoff wavelength $\lambda_c = 1.55$ μm for a silica fiber with refractive index $n_r = 1.46$. The CWV of this BBR is shown in Fig. 4(a). We assume that the required flatness of this resonator between its "ears" is ensured by the uniformity of the original fiber and the ears can be fabricated with the precision of a fraction of pm, which is achieved in SNAP technology (see, e.g., [20], where such small variations were introduced). In order to model the effect of fabrication errors, we introduce the 1 pm high Gaussian-shaped perturbation of the CWV near the right ear of this resonator (inset in Fig. 4(a) and Fig. 4(a1)). This height of this perturbation is ~ 1% of the ear height. Fig. 4(b) and its magnification in Fig. 4(b1) compare the behaviour of the intensity of the corresponding unperturbed and perturbed fundamental modes. The deviation from uniformity found from Fig. 4(b2) is ~ 2%.

In conclusion, we have determined the cutoff wavelength (effective radius) profile of a bottle microresonator which fundamental mode is exceptionally uniform along its axial length comparable with its size. The introduced microresonator, called the bat bottle resonator (BBR), may have important applications in sensing of micro/nanoparticles and assembling of quantum emitters at the microresonator surface. The BBR considered here is a simplest resonator with uniform fundamental mode distribution. Further optimization of the microresonator profile in order to maximize its mode uniformity and, simultaneously, minimize the effect of fabrication errors is critical for its potential experimental realization using the SNAP fabrication platform. We suggest that the approach similar to that developed in this Letter can be used to determine the shape of bottle microresonators having the predetermined (not necessarily uniform as considered here) axial distribution of the fundamental as well as higher order axial modes.

**Funding.** Engineering and Physical Sciences Research Council (EP/P006183/1).

**Disclosures**. The author declare no conflicts of interest.